\documentclass[9pt,conference]{IEEEtran}
\usepackage{amssymb,amsthm,amsmath,array}
\usepackage{graphicx}
\usepackage[caption=false,font=footnotesize]{subfig}
\usepackage{xspace}
\usepackage[sort&compress, numbers]{natbib}
\usepackage{stmaryrd}
\usepackage{xcolor}
\usepackage{mathtools}
\usepackage{float}
\usepackage{textcomp}

\begin{document}

\title{3D-Image Reconstruction Using MIMO-SAR FMCW Radar\\

}

\author{\IEEEauthorblockN{
        Ayush Jha\IEEEauthorrefmark{1}\IEEEauthorrefmark{5}, 
        Dhanireddy Chandrika\IEEEauthorrefmark{2}, 
        Chandra Sekhar Seelamantula\IEEEauthorrefmark{1}, and
        Chetan Singh Thakur\IEEEauthorrefmark{2}
    }
    \IEEEauthorblockA{
        \IEEEauthorrefmark{1} Department of Electrical Engineering, 
        \IEEEauthorrefmark{2} Department of Electronic Systems Engineering, 
        Indian Institute of Science, Bangalore - 560 012\\
        \IEEEauthorrefmark{5} Indian Space Research Organization, Space Applications Centre (ISRO-SAC), Ahmedabad.\\
    Email: {ayushjha, chandrikad, css, csthakur}@iisc.ac.in}
}
\maketitle
\begin{abstract}
    With the advancement of millimeter-wave radar technology, Synthetic Aperture Radar (SAR) imaging at millimeter-wave frequencies has gained significant attention in both academic research and industrial applications. However, traditional SAR imaging algorithms primarily focus on extracting two-dimensional information from detected targets, which limits their potential for 3D scene reconstruction. In this work, we demonstrate a fast time-domain reconstruction algorithm for achieving high-resolution 3D radar imaging at millimeter-wave (mmWave) frequencies. This approach leverages a combination of virtual Multiple Input Multiple Output (MIMO) Frequency Modulated Continuous Wave (FMCW) radar with the precision of SAR imaging, setting the stage for a new era of advanced radar imaging applications.
\end{abstract}


\section{Introduction}
Over the past two decades, millimeter-wave (mmWave) radars have emerged as a transformational imaging modality for various applications, including concealed weapon detection (CWD) [1], nondestructive evaluation, biomedical imaging, autonomous navigation [2], [3], and indoor mapping. The growing interest in this technology stems from the unique ability of radio frequency (RF) waves to penetrate a wide range of optically opaque and dielectric materials, such as composites, ceramics, plastics, concrete, wood, and clothing.
Compared to optical imaging systems, mmWave systems require significantly larger apertures (20–200 cm) [4]. A multiple-input multiple-output (MIMO) mmWave sensor comprises multiple transmitters and receivers. To increase the aperture for high-resolution imaging, we combine several such sensors. The hybrid MIMO-SAR concept effectively leverages the strengths of both technologies -- MIMO and SAR.\\
\indent Our goal is to exploit the wideband capabilities of mmWave sensors and the cascaded sensor configuration to enable 3D MIMO-SAR imaging with enhanced resolution and efficiency.

\section{Signal Model}
A two-dimensional (2D) synthetic antenna aperture can be synthesized by moving a cascaded MIMO radar sensor along a parallel track configuration  [5], [6], as shown in Fig.~1. We deploy a MIMO radar AWR2243 from Texas Instruments, comprising four single-chip mmWave sensors [7]. 
This configuration results in a virtual array consisting of 86 non-overlapping channels along the \(y\)-axis. The virtual elements are evenly spaced with an inter-element distance of \(\lambda/2\).

For a 3D scene characterized by the reflectivity function \(p(x, y, z)\), the primary objective in MIMO-SAR imaging is to reconstruct \(p(x, y, z)\). This is achieved by processing 5D received signal \(s(x_T, y_T, x_R, y_R, k)\), where \((x_T, y_T)\) and \((x_R, y_R)\) denote the transmitter and receiver positions, respectively, and \(k\) represents the wavenumber. The data is collected across the \(xy\)-plane from each transceiver pair, enabling detailed reflectivity recovery for the target. The system is operated in Time-Division Multiplexing Mode (TDM).

\begin{figure}[!t]
\centerline{\includegraphics[width=0.7\columnwidth]{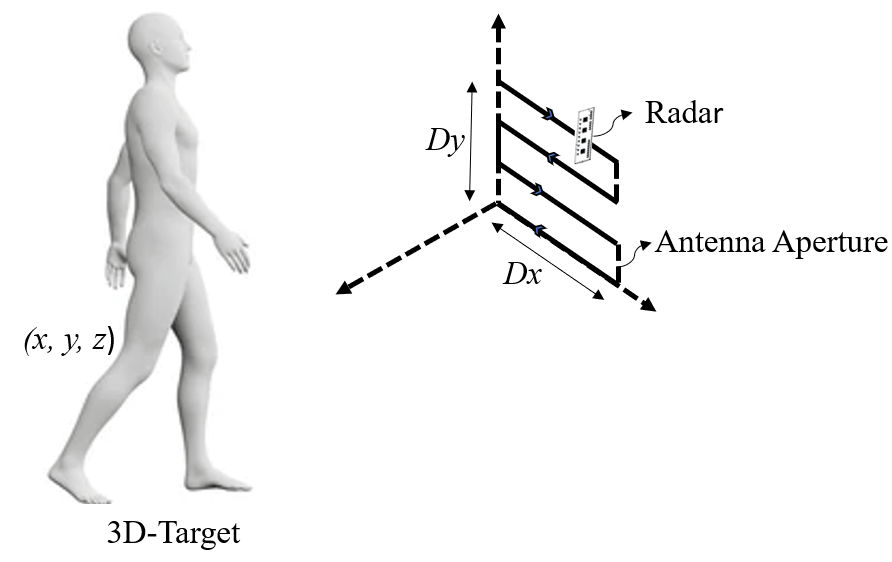}}
\caption{A 3D target is illuminated using a cascaded MIMO sensor mounted on a mechanical scanner platform with the trajectory as shown.}
\label{fig}
\end{figure}

\subsection{Algorithm for Image Reconstruction}\label{AA}
Transforming the signal into the wavenumber domain of an FMCW radar, the beat signal simplifies to
\[
s(k) = p e^{j k (R_T + R_R)}\tag{1}
\]
where \(p\) represents the reflectivity of the target, and \(k = \frac{2\pi f}{c}\) is the wavenumber associated with the signal frequency \(f = f_0 + Kt\), $K$ being the chirp rate, $R_T$ denotes the distance between the target and transmitter, and likewise $R_R$ denotes the distance between the target and the receiver. This is a compact representation of the beat signal in terms of the wavenumber, facilitating its use in radar imaging and signal processing.

Assuming a linearized scattering model, the reflectivity function \(p(x, y, z)\) of a 3D target can be used to represent the backscattered data in its effective monostatic form as 
\[
{s}(x_0, y_0, k) = \iiint p(x, y, z) e^{-j2kR} \, dx \, dy \, dz,\tag{2}
\]  
where \((x_0, y_0, z_0)\) represents the antenna location and \(R\) represents the distance between the phase center of the transceiver elements and the target point in 3D space.  

Using Weyl's decomposition of a spherical wave into a sum of plane waves [8], the term \(e^{-j2kR}\) can be approximated as follows: 
\[
e^{-j2kR} \approx \iint e^{-j[k_x (x - x_0) + k_y (y - y_0) + k_z (z - z_0)]} \, dk_x \, dk_y,
\tag{3} \]  
and substituting this into the integral for \({s}(x_0, y_0, k)\) allows the application of Fourier transform properties. The resulting spectrum of the backscattered data is expressed as:  
\[
{S}(k_x, k_y, k) = P(k_x, k_y, k_z) e^{j k_z z_0},\tag{4}
\]  
where the wavenumber components \(k_x\), \(k_y\), and \(k_z\) are related to the spatial dimensions \(x\), \(y\), and \(z\), respectively. 
We use the dispersion relation associated with the Weyl identity in Equation (4): $
4k^2 = k_x^2 + k_y^2 + k_z^2.$
This relation enables us to transform the measured beat signal from the time domain (\(t\)) to the spatial frequency domain along \(k_z\).  Since the beat signal is sampled uniformly along the \(t\) axis, the corresponding samples in \(k_z\) are nonuniform due to the nonlinear transformation. To reconstruct \(p(x, y, z)\) from Equation (4), the transformed signal \({s}(x_0, y_0, k)\) must first be interpolated onto a uniformly sampled grid in \(k_z\). To ensure the backscattered data spectrum \({S}(k_x, k_y, k)\) is uniformly sampled in the \(k\)-domain, the data is resampled to uniform positions in the \(k_z\)-domain. Finally, the reconstructed 3D image of the target can be computed as:  
$
p(x, y, z) = \mathcal{F}^{-1}_{3D}(k_x, k_y, k_z) \left[ e^{-j k_z z_0} {S}(k_x, k_y, k_z) \right]$, 
where \({S}(k_x, k_y, k_z)\) is the resampled data spectrum on a uniform \(k_z\) grid, and 
\(\mathcal{F}^{-1}_{3D}\) represents the 3D inverse Fourier transform along \(k_x\), \(k_y\), and \(k_z\) [5], [9].

\subsection{Measurements and Results}
The system shown in Fig. 2 consists of a
cascaded mmWave sensor from Texas Instruments [7], a two-axis mechanical scanner, a motor controller, and a host personal computer (PC). 
\begin{figure}[!t]
\centerline{\includegraphics[width=0.35\columnwidth]{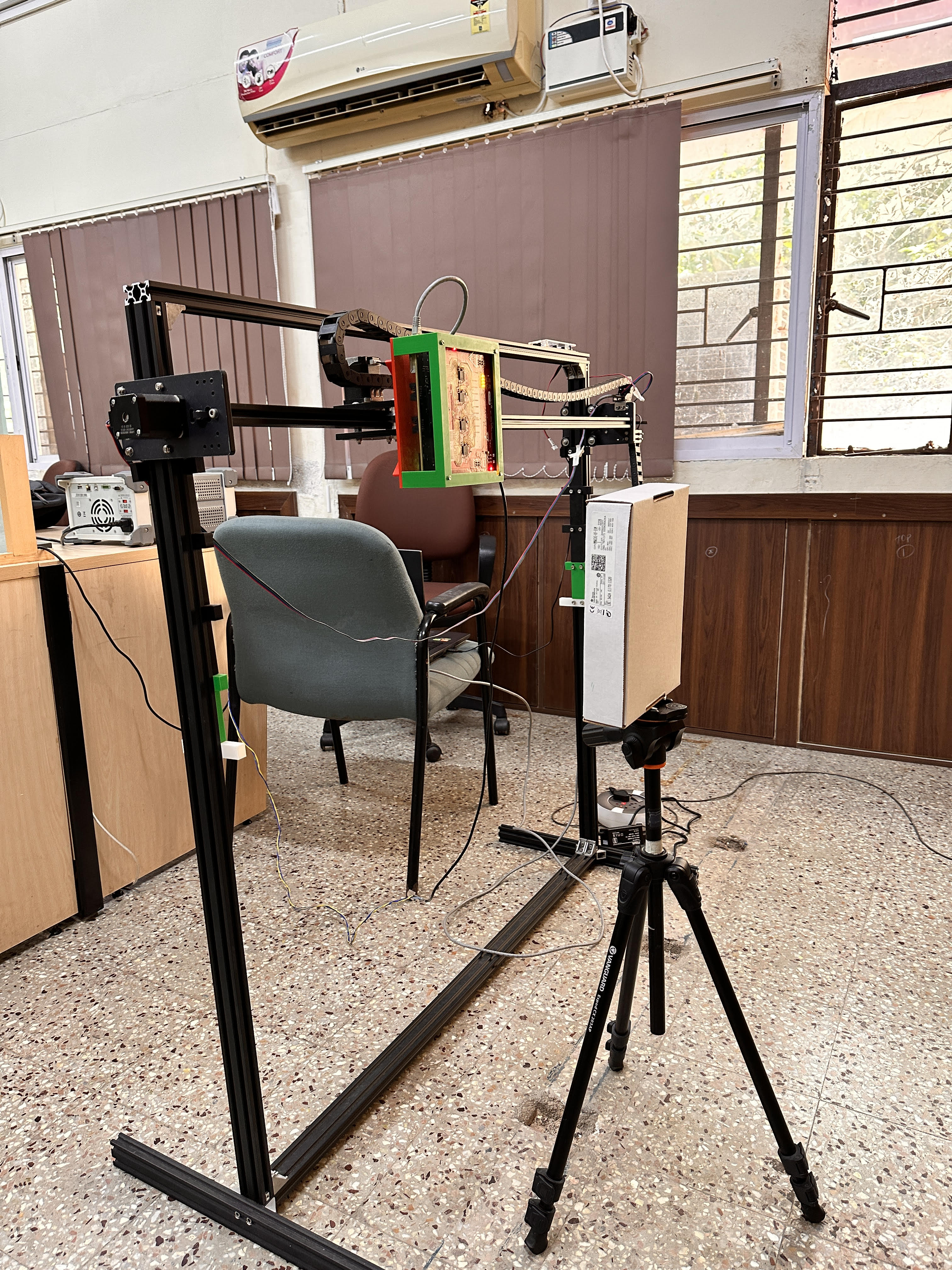}}
\caption{Measurement setup consisting of a radar cascade board and the scanning platform.}
\label{fig}
\end{figure}

To evaluate the performance of the cascaded mmWave sensor, the experiment is carried out with both uncovered and concealed targets. The FMCW waveforms used in these experiments are configured with a starting frequency of \(f_0 = 77 \, \text{GHz}\) and ending frequency of \(f_T = 80.5 \, \text{GHz}\). The signal duration is  \(T = 56 \, \mu\text{s}\), and the frequency slope is set to \(K \approx 70.295 \, \text{MHz}/\mu\text{s}\). 

In the first experimental setup, a scissor is used as the target, positioned in front of the radar as shown in Fig. 3. The target is placed at a distance of \( \approx 540 \, \text{mm}\) from the scanner. The synthetic aperture radar (SAR) scans an area of \(D_x \approx 500 \, \text{mm}\)  along the \(x\)-axis and \(D_y \approx 340 \, \text{mm}\) along the \(y\)-axis. As demonstrated in Fig. 3, the implemented algorithm successfully reconstructs the image of a scissor positioned at 54 cm  away from the radar plane.

\begin{figure}[htbp]
\centerline{\includegraphics[width=0.9\columnwidth]{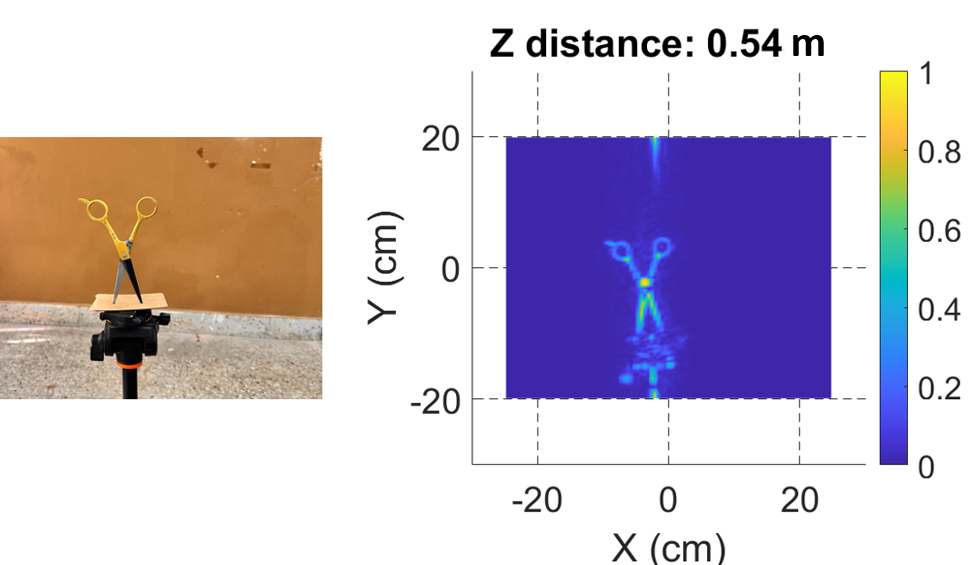}}
\caption{Optical image (left) and reconstructed image (right) using radar.}
\label{fig}
\end{figure}
To generate 3D reconstructions, a controlled experimental setup was designed as shown in Fig.~4, involving objects positioned at varying distances from the radar to evaluate its imaging capabilities. The setup includes a metallic sheet with holes of 10 mm diameter and a spanner inside a cardboard box, separated by a distance of about 80 mm. The metallic sheet and the bottom of the box are at an approximate distance of 410 mm from the radar, while the spanner was at a greater distance of  ~500 mm inside the box. Imaging frames were generated at 4 cm intervals, corresponding to the radar resolution of 4 cm along the $z$-axis.

\begin{figure}[htbp]
\centerline{\includegraphics[width=1\columnwidth]{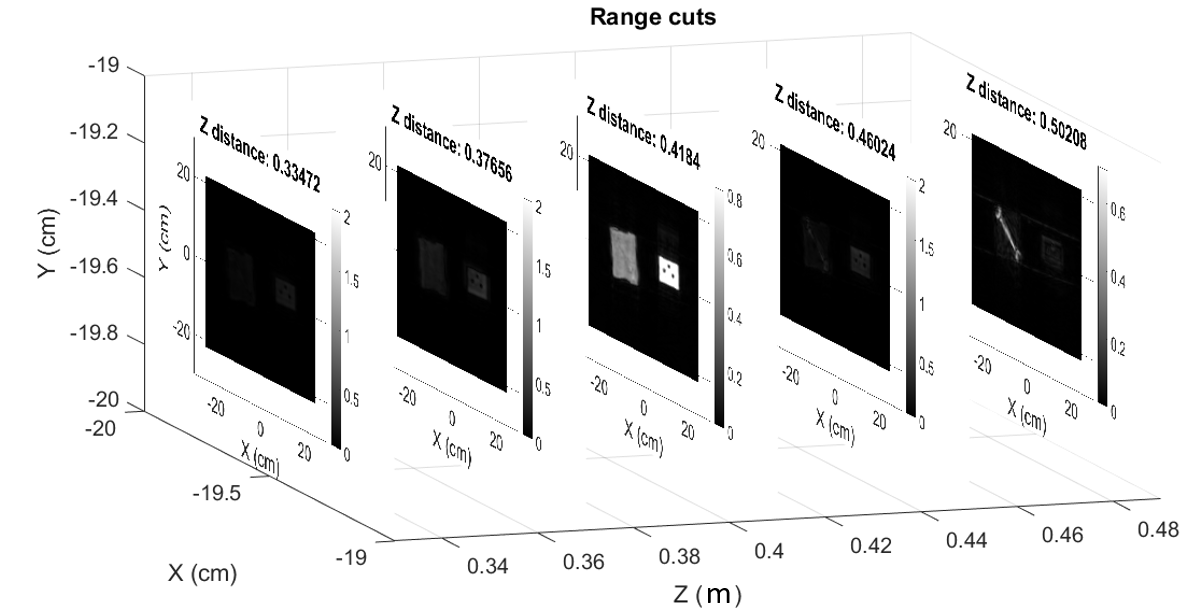}}
\caption{Experimental results of a 3-D imaging scenario where the target is a metal sheet and cardboard box with a concealed spanner. The images show different range cuts of the target.}
\label{fig}
\end{figure}
\begin{figure}[htbp]
\centerline{\includegraphics[width=0.61\columnwidth]{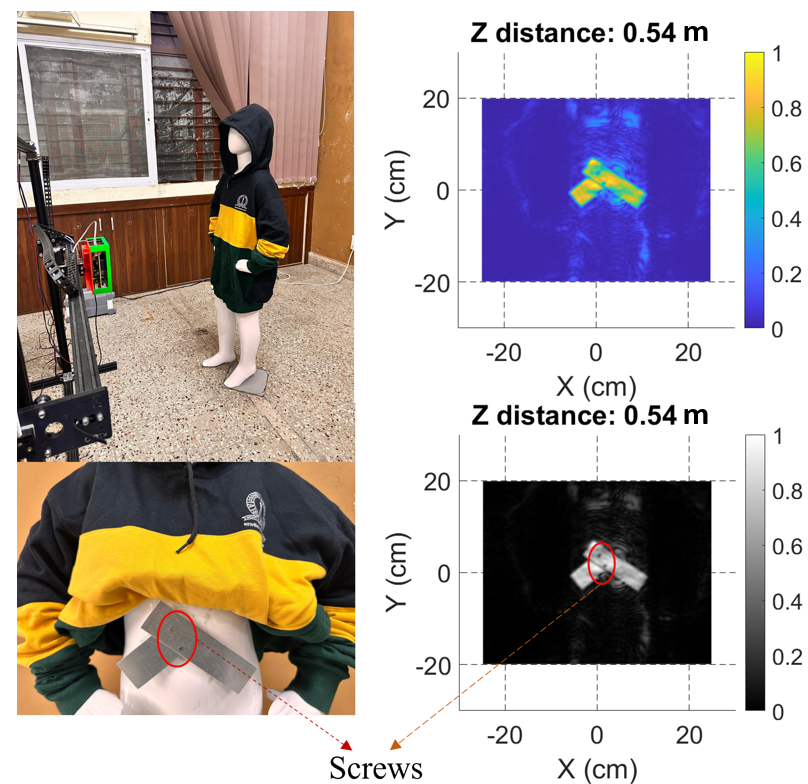}}
\caption{Experimental result demonstrating the detection of a concealed object through optically opaque material.}
\label{fig}
\end{figure}

\section{Conclusion}
The use of a FMCW cascaded radar in conjunction with the Synthetic Aperture Radar (SAR) technique was successfully demonstrated to synthesize a large antenna aperture. This approach enabled the generation of high-resolution images. Furthermore, the application of Fourier-based algorithms allowed for effective 3D image reconstruction, showcasing the potential for detailed spatial representation of target scenes. 

In addition to imaging capabilities, the experiment highlighted the ability of mmWave radar to penetrate through optically opaque materials, such as wooden enclosures. This demonstrates the effectiveness of the method in scenarios requiring non-invasive imaging through dielectric barriers, broadening the scope of applications to weapon detection, object detection for autonomous vehicles, detection of pedestrians and obstacles in low-visibility conditions, non-invasive imaging of tissues or organs, etc.

\vspace{12pt}


\newpage
\begin{thebibliography}{9}
\bibitem{b1} D. M. Sheen, D. L. Mc Makin, and T. E. Hall, “Three-dimensional
millimeter-wave imaging for concealed weapon detection,” \emph{IEEE
Trans. Microw. Theory Techn.}, vol. 49, no. 9, pp. 1581–1592,
Sep. 2001.
\bibitem{b2} J. M. Merlo and J. A. Nanzer, “A C-band fully polarimetric automotive
synthetic aperture radar,” \emph{IEEE Trans. Veh. Technol.}, vol. 71, no. 3,
pp. 2587–2600, Mar. 2022.

\bibitem{b3}  M. Steinhauer, H.-O. Ruoss, H. Irion, and W. Menzel, “Millimeter-waveradar sensor based on a transceiver array for automotive applications,”
\emph{IEEE Trans. Microw. Theory Techn.}, vol. 56, no. 2, pp. 261–269,
Feb. 2008.
\bibitem{b4} M. T. Ghasr, S. Kharkovsky, R. Bohnert, B. Hirst, and R. Zoughi, ‘‘30 GHz
linear high-resolution and rapid millimeter wave imaging system for
NDE,’’ \emph{IEEE Trans. Antennas Propag.}, vol. 61, no. 9, pp. 4733–4740,
Sep. 2013.


\bibitem{b8} M. E. Yanik, D. Wang and M. Torlak, "Development and Demonstration of MIMO-SAR mmWave Imaging Testbeds," in \emph{IEEE Access}, vol. 8, pp. 126019-126038, 2020, doi: 10.1109/ACCESS.2020.3007877. 
\bibitem{b9} L. Tang, H. Meng, X. Chen, J. Zhang, L. Lv and K. Liu, "A Novel 3D Imaging Method of FMCW MIMO-SAR," \emph{2018 China International SAR Symposium (CISS)}, Shanghai, China, 2018, pp. 1-5, doi: 10.1109/SARS.2018.8551995

\bibitem{b7} Imaging radar using cascaded mmwave sensor reference design.
[Online]. Available: http://www.ti.com/tool/TIDEP-01012

\bibitem{b8} H. Weyl, ‘‘Ausbreitung elektromagnetischer wellen über einem ebenen
leiter,’’ \emph{Ann. Phys.}, vol. 365, no. 21, pp. 481–500, 1919.
\bibitem{b9} A. V. Muppala, A. Y. Nashashibi, E. Afshari and K. Sarabandi, "Fast-Fourier Time-Domain SAR Reconstruction for Millimeter-Wave FMCW 3-D Imaging," \emph{IEEE
Trans. Microw. Theory Techn.}, vol. 72, no. 12, pp. 7028-7038, Dec. 2024, doi: 10.1109/TMTT.2024.3406938.

\end{thebibliography}
\end{document}